\documentclass{aa}

\usepackage{natbib}
\usepackage{amssymb,amsmath,wasysym}
\usepackage{txfonts}
\usepackage[colorlinks=true, linkcolor=blue, urlcolor=red, citecolor=blue]{hyperref}
\usepackage{cleveref}
\usepackage{afterpage}
\usepackage{local_macros}

\begin{document}
\title{Toward a new paradigm for Type II migration}
\titlerunning{Type II migration}

\author{C. M. T. Robert \inst{1}
  \and A. Crida\inst{1,2}
  \and E. Lega\inst{1}
  \and H. Méheut\inst{1}
  \and A. Morbidelli\inst{1}
}

\institute{Universit\'e C\^ote d'Azur, Observatoire de la C\^ote d'Azur, CNRS, Laboratoire Lagrange, Bd de l’Observatoire, CS 34229, 06304 Nice cedex 4, \textsc{France}
  \and
  Institut Universitaire de France, 103 Boulevard Saint-Michel, 75005 Paris, \textsc{France}
  \\\email{clement.robert@oca.eu}
}

\date{\today}

\abstract
{ Giant planets open gaps in their protoplanetary and subsequently
suffer so-called type~II migration. Schematically, planets are thought
to be tightly locked within their surrounding disks, and forced to
follow the viscous advection of gas onto the central star.
This fundamental principle however has recently been questioned, as
migrating planets were shown to decouple from the gas' radial drift.
}
{ In this framework, we question whether the traditionally used linear
scaling of migration rate of a giant planet with the disk's viscosity
still holds.
Additionally, we assess the role of orbit-crossing material as part of
the decoupling mechanism.  }
{ We have performed 2D ($r,\theta$) numerical simulations of point-mass
planets embedded in locally isothermal $\alpha$-disks in steady-state
accretion, with various values of $\alpha$.
Arbitrary planetary accretion rates were used as a means to diminish or
nullify orbit-crossing flows.  }
{ We confirm that the migration rate of a gap-opening planet is indeed
proportional to the disk's viscosity, but is not equal to the gas
drift speed in the unperturbed disk.
We show that the role of gap-crossing flows is in fact negligible.}
{ From these observations, we propose a new paradigm for type~II
migration\,: a giant planet feels a torque from the disk that promotes
its migration, while the gap profile relative to the planet is
restored on a viscous timescale, thus limiting the planet migration
rate to be proportional to the disk's viscosity.
Hence, in disks with low viscosity in the planet region, type~II
migration should still be very slow.  }
\keywords{protoplanetary disks -- planet-disk interactions -- planets and satellites: formation}
\maketitle

\section{Introduction}

Planetary migration is a key ingredient to understand the architecture
of planetary systems.
This radial displacement of planets is due to their gravitational
interaction with the protoplanetary disk.
These disks surround most young stars, and have a lifetime of a few
million years.
Planetary migration leads to significant changes in the semi-major
axis of all planets \citep[see][for a recent review]{baruteau+2014}
and carves the structure of planetary systems.

Migration of planets has been extensively studied in recent decades.
Small mass planets, for which the response of the disk can be
considered linear, do not perturb the density profile of the disk, and
are in a regime called type~I migration.
Giant planets, however, are massive enough to modify the disk radial
density profile.
They deplete the region around their orbit and create a gap
\citep{lin-papaloizou1986a}, separating the inner from the outer disk.
Once a gap is open, the planet is repelled inwards by the outer disk
and outwards by the inner disk.
The position of the planet within the gap adjusts so that the torques
from the inner and out disks cancel out.
However, as the disk spreads viscously and the gas accretes onto the
central star, the gap, as well as the embedded planet that carved it,
are carried with it.
This is the classical scheme of the so-called type~II migration
\citep{lin-papaloizou1986b}, responsible for inward motion of giant
planets.
In this scheme, the planet does not migrate with respect to the
gas, but together with the gas, and acts as a gas-proof barrier
between the parts of the disk.

This standard scheme of type~II migration, where a planet follows
exactly the viscous accretion speed of the gas has been questioned by
several works.
\citet{quillen+2004} note that if the inertia of the planet is much
larger than that of the gas originally present in the gap, the disk
has a hard time moving the planet, and the migration is slower than
the viscous speed.
\citet{crida-morbidelli2007} add that the corotation torque, exerted
on the planet by the gas still present in the gap, may play a role,
especially in regions where the background density profile is steep
(which promotes a high corotation torque).
This could slightly decouple the planet from the gas evolution, but
this process relies on a non-empty gap, hence it could be seen as a
situation where perfect type~II migration is not expected anyway.

Furthermore, \citet{hasegawa-ida2013} remark that the assumption
that a planet be locked in its gap had no solid physical
ground.
\citet{lubow-dangelo2006} and \citet{duffell+2014} show that, in
simulations, gas is able to cross the gap during planetary
migration. \citet{duermann-kley2015} (DK15 hereafter) explicitly
question the idea that the planet stays in equilibrium in the middle
of the gap.
They suggest that when the gas reaches an equilibrium gap profile,
such that the torques from the viscosity, the pressure, and the
planetary gravity balance on each side \citep{crida+2006}, the planet
does not necessarily feel a zero torque.
Hence it moves to a different position inside the gap.
The motion of the planet then forces the gas to re-adjust the
equilibrium profile, by passing through the planet's orbit from the
inner to the outer part of the disk.
Because this transfer of gas is due to the planet-gas interaction and
not to the disk's viscous spreading, the evolution of the planet
becomes unlocked from the disk's viscous evolution.

The author of DK15 find that the migration speed is independent of the disk's drift
speed and mainly depends on parameters such as the mass of the disk or
that of the planet.
Furthermore, \cite{duermann-kley2017} (DK17 hereafter) showed that
planetary accretion, in some cases, is able to cut the gas flow across
the planet's orbit.
In these cases the gap acts as a barrier between the inner and the
outer disks and classical type~II migration regime could be
reestablished.
However the authors noticed that even in these cases the migration rate
can differ from classical type~II migration rate.
Both studies (DK15 and DK17) considered a classical $\alpha$-viscosity
disk and focused on the dependence of migration speed on parameters
like the disk mass and the planetary mass for a fixed viscosity value.

These studies show that gas is crossing the gap and therefore the
migration speed seems to be independent on the disk's drift.
Although the authors already provided evidence that
  migration speeds depend on viscosity, a direct comparison of this
  dependence with the fundamental assumptions of classical type~II
  remains to be conducted.
  This is precisely the aim of the present paper.

In fact, though it is now admitted that giant planets' migration
does not follow classical type~II migration, it is very important to
check whether some scaling of migration rate with viscosity is still
preserved.
Precisely, the existence of giant planets at orbits larger than $1$
a.u. in semi-major axis (so called warm Jupiters) is difficult to
explain in the usual paradigm of viscously accreting disks unless
considering low viscosity and confirming that migration speed scales
with viscosity.
Low viscosity is admitted in the central part of the disks, the
so-called dead-zone, where magneto-rotational instability
\citep{BalbusHawley91} does not operate.
Moreover, recent studies \citep{BaiStone2013,Bai2016,Suzuki+2016} have
shown that the disk's structure can be very different from that of a
viscously accreting disk.
Mass accretion onto the star could be ensured by magnetically driven
winds providing angular momentum removal.
These disks would have very low viscosity.
It is beyond the purpose of this paper to model this kind of disks,
however these studies motivate the investigation of the migration of
giant planets in low viscosity disks.
We will also investigate the role of gap crossing flows on migration.

The paper is organized as follows.
In \Cref{sec:setup}, we present the physical model and the numerical
scheme, while the setup of initial conditions is reported in the
appendix.
In \Cref{sec:viscosity}, we study the scaling of migration speed with
viscosity by performing numerical simulations of Jupiter-mass planets
in disks with controlled inflow rates and various viscosities, but
same gas surface density.
In \Cref{sec:accretion}, the accretion of gas by the planets is
modeled by removing, with various efficiencies, the gas entering the
Hill sphere\,; it allows us to find the influence of planetary accretion
on migration, and to quantify how important is cutting the gas flow
across the gap.
Finally, our findings are put summarized in \Cref{sec:conclu}, where
we propose a new, consistent paradigm to explain giant planet
migration.

\section{Physical setup}
\label{sec:setup}

We consider a stationary accreting disk in which a planet is
introduced.
In this section, we present our background disk model the numerical
code used and the prescription for planetary accretion.

\subsection{Units and notations}
We describe our 2D-disk in polar coordinates $(r,\,\theta)$, centered
onto the host star.
A subscript ``$_0$'' denotes values defined at a reference radius
$r_0=1$\,.  Our time unit, hereafter called an \emph{orbit}, is $t_0 =
2\pi \sqrt{\frac{r_0^3}{Gm_*}} = \frac{2\pi}{\Omega_0}$, where $m_*$
is mass of the central star.
The planet mass is defined as $m_p = qm_* = 10^{-3}m_*$\,, and its
initial semi-major axis as $r_p(t=0)=r_0=1$\,.

\subsection{Accretion disk model} 
\label{ssec:model}

Because we are interested in comparing the radial drift of the
embedded planet to that of the unperturbed disk, the easiest scheme is
to setup a steady-state accreting disk.
The local accretion rate is defined as
\begin{equation}\label{eq:def_mdot}
  \dot{M}(r) = - 2\pi r v_r \Sigma\;,
\end{equation}
hence, an accreting disk displays $v_r<0$ and $\dot{M}>0$, where $r$
is the distance to the central star, $\Sigma$ the gas surface density,
and $v_r$ the gas radial velocity.
We use a non-flared disk scale-height $H(r) = hr$, $h=0.05$ being the uniform aspect ratio,
as well as an $\alpha$-viscosity model \citep{shakura1973} $\nu=\alpha H^2 \Omega$\,,
and a power law density profile, $\Sigma(r)=\Sigma_0(r/r_0)^{-s}$\,.
Within those prescriptions, we obtain the radial velocity for a uniform (hence steady) accretion
rate as\footnote{ in all generality, for a $\beta \neq 0$ flared disk
  $H(r)=hr(r/r_0)^\beta$, the result is changed by replacing $(1-s)$
  by $(1-s+2\beta)$\,.}
\begin{equation}\label{eq:steady_vr}
  v_r = -3(1-s)\frac{\nu(r)}{r}\;.
\end{equation}
Combining \Cref{eq:steady_vr,eq:def_mdot} gives $\dot{M} = \dot{M}_\mathrm{ref} (1-s)\alpha (r/r_0)^{1/2-s}$, where $\dot{M}_\mathrm{ref} = 6\pi h^2 r_0^2 \Omega_0 \Sigma_0 $\,.
Therefore, the disk model is completely defined by fixing $\alpha$ and
$\dot{M}$.
This leads to
\begin{equation}\label{eq:defsigma}
  s = \frac{1}{2}
  \,,\,
  \Sigma_0 = \frac{\dot{M}}{3\pi \alpha h^2 r_0^2\Omega_0}\,.
\end{equation}

Unless specifically stated, all our simulations share this set-up,
following \Cref{eq:steady_vr,eq:defsigma}.
A summary of different values used in this study may be found in
\Cref{tab:values}.
In \Cref{ssec:bc}, we exhibit boundary conditions compatible with this physically stable initial state, and in \Cref{app:bcs}, we explain how they affect the disk in the presence of a planet.

\subsection{Hydro-thermodynamics}

The disk evolves along the Navier-Stokes equations\,:
\begin{align}
  \label{eq:navier-stokes}
  &\left( \frac{\partial .}{\partial t} + \vect{v} . \Nabla \right)
  \vect{v}
  =
  \nu \Delta \vect{v} + \nabla \Phi_G -\frac{\Nabla P}{\Sigma}\ ,\\
  \label{eq:mass_conservation}
  &  \left( \frac{\partial .}{\partial t} + \vect{v} . \Nabla \right)\Sigma
  + \Sigma (\Nabla . \vect{v}) = 0\ ,
\end{align}
where $\vect{v}$ is the gas velocity, $\nu$ is the $\alpha$-viscosity,
$\Phi_G$ is the total gravitational potential yielded by the central
star and the planet, and $P$ is the pressure.
Assuming a cooling time much shorter than the orbital period
$2\pi\Omega^{-1} = 2\pi (r/r_0)^{-3/2} \Omega_0^{-1}$, the equation
system is closed with a locally isothermal equation of state\,:
\begin{equation}
  P = {c_s}^2(r)\ \Sigma = (H \Omega)^2 \Sigma\ ,
\end{equation}
where $c_s$ is the sound speed and $H(r) = h r$ is the disk scale height.

\subsection{Numerical code}

Our experiments were conducted using the 2D hydrodynamic grid code
\texttt{Fargo 2D} \citep{masset2000}.
This code solves \cref{eq:navier-stokes,eq:mass_conservation} using a
finite difference multistep procedure.
The fluid advection step is solved using a Van-Leer method
\citep{van_leer1977}.
The \texttt{Fargo} algorithm \citep{masset2000} is specifically suited for
keplerian rotation where the traditional Courant-Friedrichs-Lewy (CFL)
condition \citep{courant+1928} provides very small time-steps due to
fast orbital motion at the inner boundary of the numerical domain.
In the \texttt{Fargo} algorithm the time-step is limited by the perturbed
density arising from differential rotation.

Our solver was set with a Courant parameter \citep{courant+1928}
of $0.4$.
The contribution $\Phi_p$ of the planet to the gravitational potential is
smoothed as
\begin{equation}
  \Phi_p = \frac{G m_p}{\sqrt{d^2+\left(\frac{3}{5}R_H\right)^2}}\ ,
\end{equation}
where $d$ is the local distance to the planet, and $R_H =
r_p(q/3)^{1/3}$ is the planet's Hill radius.
As our model does not include the disk's self-gravity, we exclude the
material contained in the Hill region of a planet in the computation
of the torque acting on it.
To this end, we use the tapering function described by
\cite{crida+2008} (eq. (5)), with $p=0.6\,$.
This is done to avoid artificial ``braking'' in the planet's migration
due to the fact that this circum-planetary material does not feel any
gravitational torque from the rest of the disk
(see \cite{crida+2009}).

\subsection{Planetary accretion recipe}

Accretion onto the planet is handled following \cite{kley1999}'s
recipe.
At each time step, in every cell located within the Hill radius $R_H$
of the planet, a fraction of the gas is removed.
This fraction is given by $K_p\,f(d)\,\delta t$, where $d$ is the
distance to the planet, $\delta t$ is the time step, and $K_p$ is an
arbitrary accretion efficiency parameter. $K_p$ is typically $\sim 1$,
and constrained such that $0\leqslant K_p\,\delta t < 1$.
In our simulations $\delta t \simeq 10^{-3}$.
We use the smooth function $f(d)$ proposed by
\citet{crida-bitsch-raibaldi2016}\,:
\begin{equation}
f(d) = \left\{\begin{array}{lll}
1 & {\rm if} & d \leqslant 0.3\,R_H\ ,\\
\cos^2\left(\pi \left(\frac{d}{R_H} -0.3\right)\right) & {\rm if} & 0.3\,R_H<d<0.8\,R_H\ ,\\
0                                                      & {\rm if} & 0.8\,R_H\leqslant d\ .
\end{array}\right.
\end{equation}
For a uniform density around the planet, this function conserves the
planetary accretion rate with respect to the original step function
proposed by \citet{kley1999}.
It furthermore provides better accuracy when the gas is growing scarce
in the planet's vicinity, without the need for a higher resolution.

This mass taken away from the disk should fall onto the planet, with
corresponding momentum.
However, here we remove this gas (and its momentum) from the
simulation, in order to compare the migration rates of planets of
equal, fixed masses.
By doing so, we highlight the influence of the gas flow on the
migration of the planet.
This is reasonable as we are interested in the migration rate of a
planet given its mass and not in the planet's mass accretion rate nor
in the final position of an evolving planet.

\subsection{Initial conditions for migration}

We obtain our initial conditions after two stages, respectively coined
\emph{introduction} and \emph{relaxation}, lasting $T_\mathrm{intro} =
1000$ and $T_\mathrm{relax} = 4000$ orbits, respectively.
During \emph{introduction} and \emph{relaxation}, the planet is held at a
constant semi-major axis $r_p=1$.

During \emph{introduction}, the planet mass is slowly increased from
$0$ to $qm_*$ over $T_\mathrm{intro}$, along a smooth function of time
\begin{equation}
  m_p (t) =  q m_* \sin^2\left(\frac{\pi}{2} \frac{t}{T_\mathrm{intro}}\right)\ .
\end{equation}
Classically, a gap-opening planet quickly expels the gas from its
horseshoe region.  This gas tends to accumulate at the gap's edges
then to slowly spread at a viscous rate.
To avoid this slow phase, we allow the planet to remove gas instead of
scattering it, using the planetary accretion recipe described above.
We use a smoothly decreasing planetary accretion efficiency
\begin{equation}\label{eq:accretion}
  K_p(t) = K_p^0\cos^2\left(\frac{\pi}{2} \frac{t}{T_\mathrm{intro}}\right)\ ,
\end{equation}
with $K_p^0=1$ in our simulations.
As illustrated by \citet{crida-bitsch2017}, this helps
with the gap opening process, and overall saves computational time.

During \emph{relaxation}, we let the system evolve to a near-to-steady
state.
Convergence is considered achieved when both the mass flow
$\dot{M}(r)$ and the gravitational torque $\Gamma_\mathrm{tot}$
exerted on the planet reach constant values with respect to time.
See \Cref{app:ics}.
Hereafter, we call the time origin $t=0$, the \emph{release date},
from which the planet is allowed to migrate.

\subsection{Boundary conditions}
\label{ssec:bc}
The choice of boundary conditions is a priori of non- negligible 
importance, and can significantly affect the numerical
  steady-state the disk relaxes into.
Our main concern in choosing appropriate boundaries is that the disk
should be left unperturbed far from the planet's orbital radius.
In order to achieve this, we make the further distinction between the
domain of interest, in other words, the radial vicinity of the planet, and the
broad edges of the simulation domain.
A smooth transition is used from the planet's region of influence to
the unperturbed initial state at edges of the simulation domain.
This is done through a wave-killing-like algorithm
\citep{Val-Borro2006} where at each time step and in dedicated regions\footnote{
  extending over $r\in[r_\mathrm{min},\, 1.25r_\mathrm{min}]$, and $r\in[0.8r_\mathrm{max},\, r_\mathrm{max}]$ respectively, $r_\mathrm{min}$ and $r_\mathrm{max}$ being the limits of the grid.},
perturbations in $\Sigma, v_r, v_\theta$
with respect to the initial state are damped out.
This design choice is justified in more extensive details in
\Cref{app:bcs}.

\subsection{Models parameters}

The simulation domain spans over $r\in [0.1,\,3.0]$.
We have chosen an inner boundary at $r=0.1$, though this choice is
very expensive in computational time.
The advantage is the possibility to study migration of a planet
initially located at $r=1$ on a large orbital domain, going down in
some cases to $r=0.4$ with no impact of the inner boundary on
migration.
An arithmetic radial-spacing of grid cells is used with a resolution of
$(n_r,\, n_\theta) =(248,\,628)$.

In this paper we aim at testing the scaling of the migration rate with
the viscosity and therefore we consider different values of $\alpha$
in the interval $[3\, 10^{-3}: 3\, 10^{-4}]$.
\Cref{tab:values} describes in code units our models parametrization.
Values are chosen so that $\Sigma_0=5.3\, 10^{-4}$ code units in all
simulations.
This value is low enough that the co-orbital mass deficit can not
exceed $m_p$, to avoid the runaway type~III migration regime
\citep{masset-papaloizou2003}, but large enough that the disk is still
able to push the planet efficiently\footnote{We notice that, for
  $\Sigma_0 r_p^2/m_p=0.53$\,, DK15 obtained a migration speed larger
  than the disk radial velocity $v_r$ (see Fig. 15 in their paper).}.
For $m_* = m_{\astrosun}$ (hence $m_p = m_J$), and $r_0 =
5\,\mathrm{a.u.}$\,, model {\tt A1} physically translates into and
$\dot{M} = 2\cdot10^{-8}m_{\astrosun}\mathrm{yr}^{-1}$\,.

\begin{table}
  \centering
  \begin{tabular}{c|c|c|}
    Simulation name &  $\alpha$  &  $\dot{M}/(m_*t_0^{-1})$ \\  
    \hline
     {\tt A1}  & $3\; 10^{-3}$ & $3.78\; 10^{-8}$ \\
     {\tt A2}  & $1\; 10^{-3}$ & $1.25\; 10^{-8}$ \\
     {\tt A3}  & $7\; 10^{-4}$ & $8.80\; 10^{-9}$ \\
     {\tt A4}  & $5\; 10^{-4}$ & $6.29\; 10^{-9}$ \\
     {\tt A5}  & $3\; 10^{-4}$ & $3.78\; 10^{-9}$\\
     {\tt B1}  & $3\; 10^{-3}$ & $0$  \\
    
  \end{tabular}
  \caption{Models parameters. In all the cases, the disk mass, or
    equivalently its surface density $\Sigma_0$ is held constant to
    $5.3\,10^{-4}$. }
  \label{tab:values}
\end{table}

\begin{figure}[!ht]
    \includegraphics[width=\hsize]{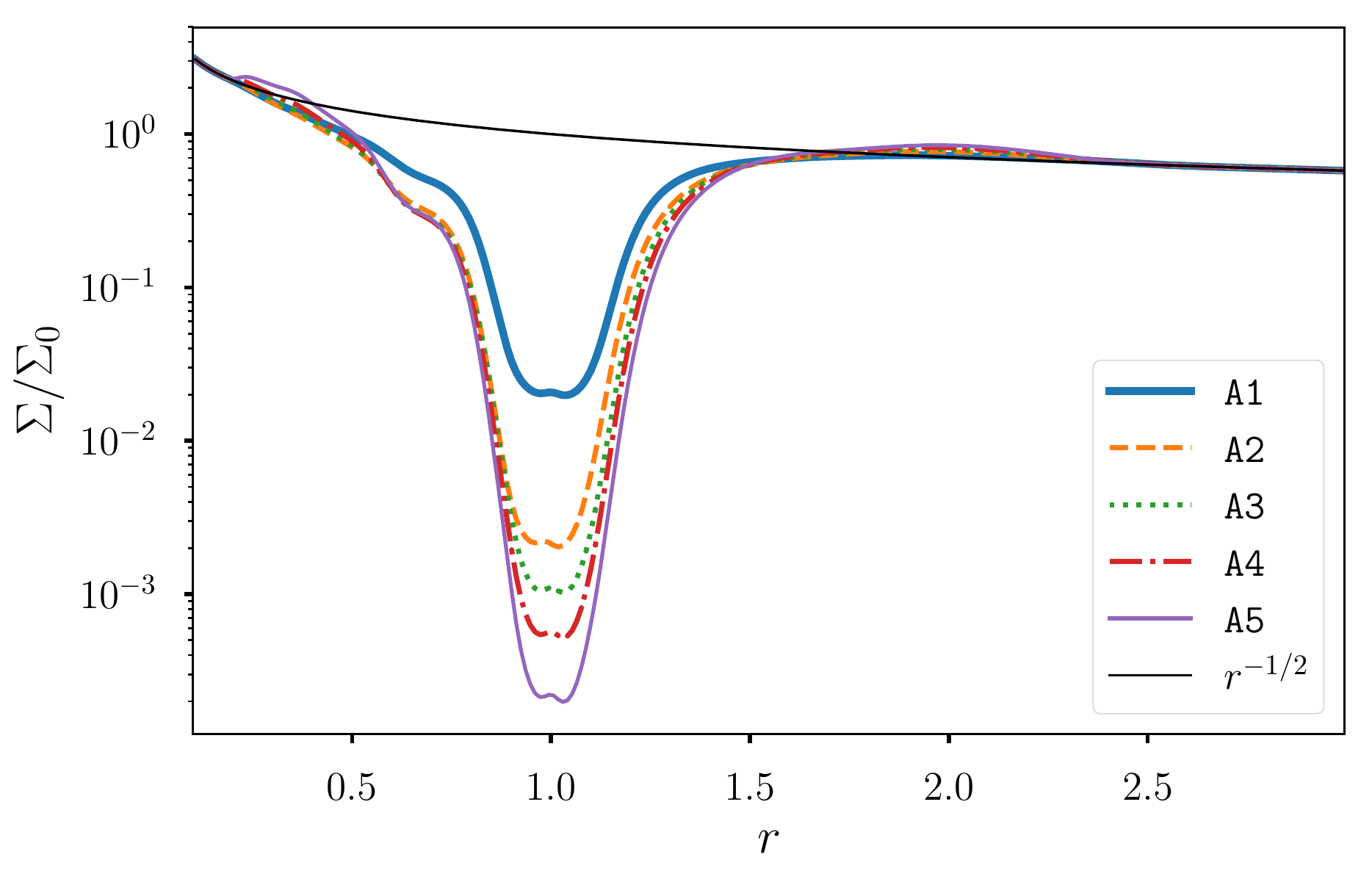}
    \caption{Azimuthally averaged density profiles at $t=0$, when the
      planet is being released.  Black thin solid curve corresponds to
      the power law profile used at initialization.}
    \label{fig:init_density}
\end{figure}

\Cref{fig:init_density} shows the initial density profiles at the end
of the \emph{relaxation} phase at $t=0$ for simulations {\tt A1}, {\tt
  A2}, {\tt A5}.
As expected gaps are deeper and wider at lower viscosities.

\section{Influence of viscosity on the migration rate}
\label{sec:viscosity}

DK15 and DK17 demonstrated that the torque acting on a gap-opening
planet is primarily dependent on the disk's mass rather than its
accretion rate $\dot{M}(r)\propto \nu(r)$.
Although they showed that viscosity still affected the torque, it remained to be clarify how viscosity's role compares to the initial assumption of classical type~II.
A direct comparison with type~II is the topic of this section.

\begin{figure}
  \includegraphics[width=\hsize]{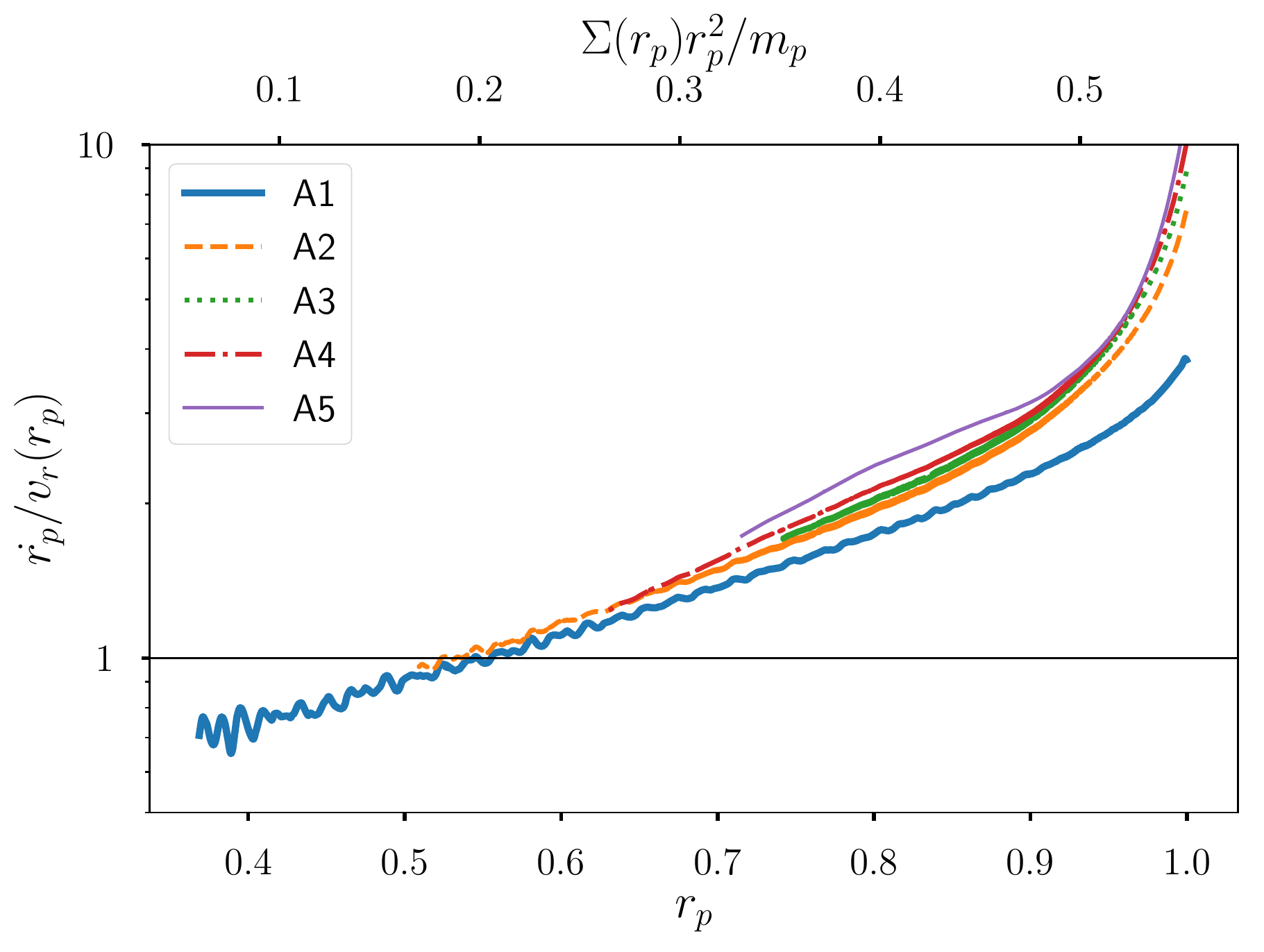}
  \caption{Normalized migration speed as a function of the semi-major
     axis for sets {\tt A1} to {\tt A5} with $K_p=0$.  The normalized
     local disk mass $\Sigma(r_p)r_p^2/m_p$ is indicated as a secondary
     graduation for the $x$ axis.
     }
  \label{fig:speeds}
\end{figure}

``Classical'' type~II migration rate is given by $\dot{r_p} =
v_r$, where $v_r$ is the viscous speed of the unperturbed disk
\Cref{eq:steady_vr}\,; i.e. the planet is assumed to migrate with the
drift rate of the gas.
We show in \Cref{fig:speeds} the migration speed measured in
simulations {\tt A1} to {\tt A5} plotted as a function of the
semi-major axis of the planet.
In all these runs, no planetary accretion is used.
The migration speed was normalized in all simulations to
$v_r=-\frac32\frac{\nu}{r}$ so that scaling of the
migration with $v_r$ and/or with viscosity is apparent.
Because planets do not migrate at identical speeds in different runs,
note that a given semi-major axis corresponds to different dates\,;
for instance, {\tt A1}'s planet reaches $r_p=0.8$ at $t\simeq1150$, to
be compared with $t\simeq8000$ in case {\tt A5}.
Some observations can be drawn from \Cref{fig:speeds}\,:

\begin{itemize}
\item As pointed out by \cite{duffell+2014}, because the planet was
  first artificially maintained on a circular orbit, the corresponding
  gas distribution at release date is \emph{de facto} inconsistent
  with a migrating perturber. Indeed, within this method, we obtain
  torques inducing migration timescales much shorter than the viscous
  spreading timescale of gap edges\,; this is a known
  effect in type~II studies. It follows that the transitional stage
  immediately following release ($r_p\approx 1$) ought to be discarded
  from our analysis.
\item Following this transition, normalized migration tracks converge,
  demonstrating that steady migration rates scale linearly with the
  viscosity.
\item Because migration speeds scale with viscosity, convergence is
  reached on timescales $\propto \tau_\nu = r_p^2/\nu$. Therefore,
  very long integration times are required at the lower viscosities.
  In fact, properly comparing situations at different values of
  $\alpha$ requires snapshots with similar $r_p$, i.e. similar
  $t/\tau_\nu$ (and similar local disk mass, see below).  In
  particular, the previously mentioned transitional stage corresponds
  to $0.9\lesssim r_p<1$ for every viscosity.
\item At odds with the classical speed expected for type~II migration,
  the migration rate decreases with the semi-major axis and migration
  becomes slower than classical type~II for $r_p\lesssim 0.54$.
\end{itemize}
This last result is in agreement with DK15 who showed that
migration rate decreases with decreasing local disk's mass $\Sigma_p r_p^2$ ,
$\Sigma_p$ being the unperturbed surface density at planet position
$r_p$.
The top horizontal axis of \Cref{fig:speeds} shows the local disk mass
divided by the mass of the planet.
The migration speed is equal to $v_r$ when this mass ratio is $\simeq
0.2$, in remarkable agreement with DK15 (Fig.15. therein).

\begin{figure}
  \includegraphics[width=\hsize]{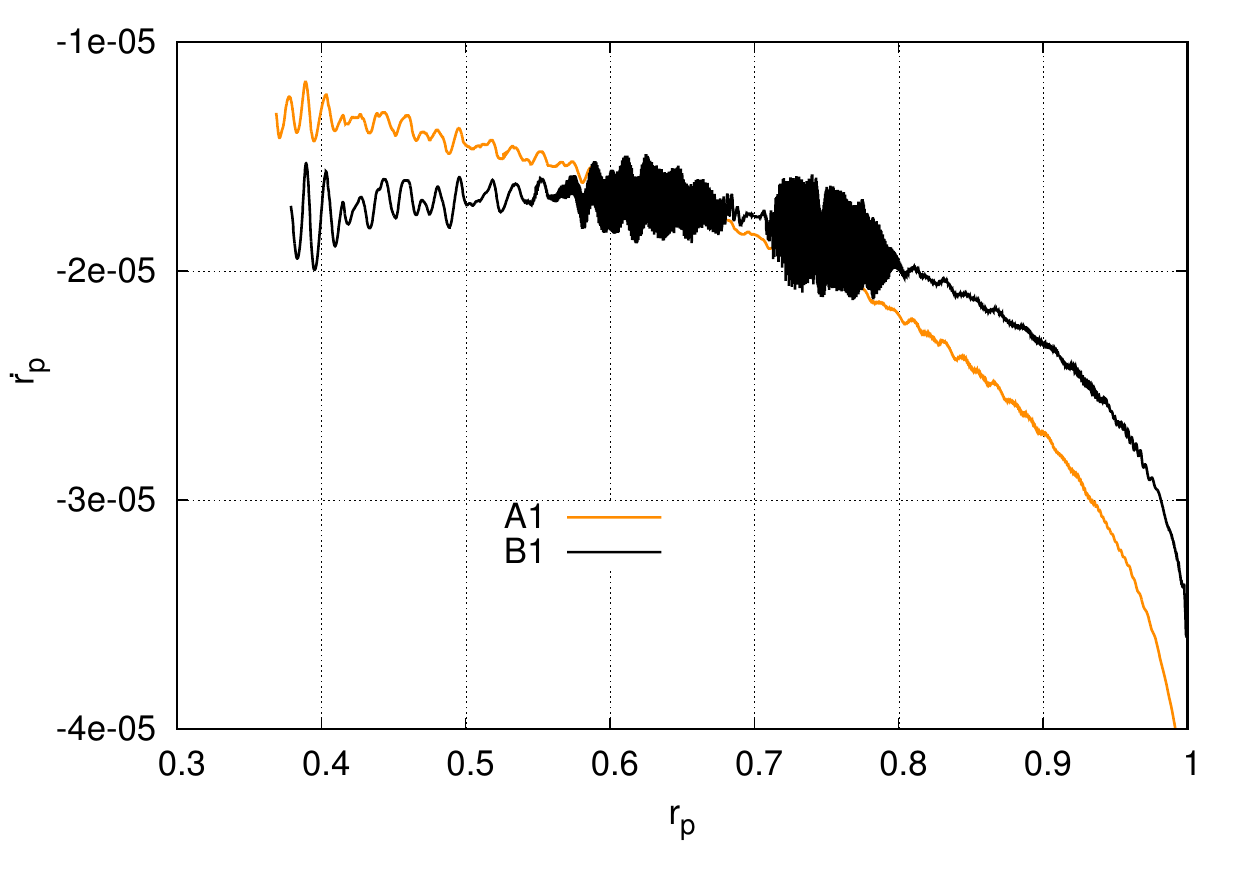}
  \caption{Migration speed as a function of the semi-major axis in
     two simulations with $\alpha=0.003$ and same surface density.
     Orange curve\,: case {\tt A1} ($s=1/2$)\,; same curve as
     \Cref{fig:speeds} (not normalized here). Black curve\,: case {\tt B1}.}
  \label{fig:equil}
\end{figure}

Besides these observations, \Cref{fig:speeds} reveals a puzzling fact
for the question we are interested in\,: the migration speed of a gap
opening planet is proportional to the viscosity of the disk, but not
equal to the radial drift of the gas.
The fact that the planet migrates slower than the gas when the disk
mass is low is not a surprise, but the reason for a faster migration
remains unclear still (although already found by previous studies).
To further inquire this possibility, we ran an additional simulation,
{\tt B1}, in a static unperturbed disk, i.e. where \Cref{eq:defsigma}
is changed to $s=1$ so that $v_r=\dot{M}=0$, (\Cref{eq:steady_vr}),
and $\Sigma_0$ is arbitrary, hence kept to {\tt A1}'s value.
The migration speed found is displayed in \cref{fig:equil} as the
black curve, where it can be compared to the case $s=1/2$ (orange
curve, same data as \Cref{fig:speeds}).
Very clearly (and surprisingly), the unperturbed radial velocity of
the gas has little influence on the migration speed of the planet.

What pushes the planet is \emph{not} the radial drift of the gas onto
the star.
This supports DK15's claim that the planet inside its gap is not
necessarily at equilibrium with the gas, when the gas profile is
itself at equilibrium with the planet.
Hence, the planet can feel a torque, which drives its migration, even
if the unperturbed disk has no radial drift.
This is already a change of paradigm for type~II migration.
Why this torque should be proportional to the viscosity, however, is
unclear at this point.
This suggests that the picture may be more complicated than the one
just described, as we will see later.

For the planet to migrate faster than the radial velocity of the
unperturbed gas.  one would naturally expect that it migrates with
respect to the gas.
In this picture, gas should cross the planet's orbit from the inner to
the outer disk to sustain migration.
The following section is dedicated to studying the influence of
planetary accretion on such mass exchanges and its resulting effects
on migration.

\section{Effect(s) of planetary accretion}
\label{sec:accretion}
Here we test how introducing planetary accretion into our model
affects the gas flow through the planet's orbit, and how the migration
rate is changed in turn.
Accretion's efficiency is parameterized by the dimensionless
number $K_p$, which we vary as $K_p \in \{0,\, 0.2,\, 1.0,\, 5.0\}$,
using run {\tt A2} as the reference case $K_p=0$.
The initial state in all runs is identical to that of {\tt A2}.
Unlike the time-dependent accretion efficiency previously described,
here we do not follow \cref{eq:accretion}, and $K_p$ is instantly
switched to a non-zero value at $t=0$, when the planet is released.

In order to preserve comparability, we emphasize the importance of
keeping the global structure of the disk as unperturbed as can be,
despite the fact we are adding a sink point to the hydrodynamical
model.
Our boundary conditions ensure that the disk's profile stays unchanged
far away from the planet.
Hence we except all changes due to $K_p$ to stay local to the planet's
vicinity.

\subsection{Material exchange between reservoirs}
\begin{figure*}
  \centering
  \includegraphics[width=\hsize]{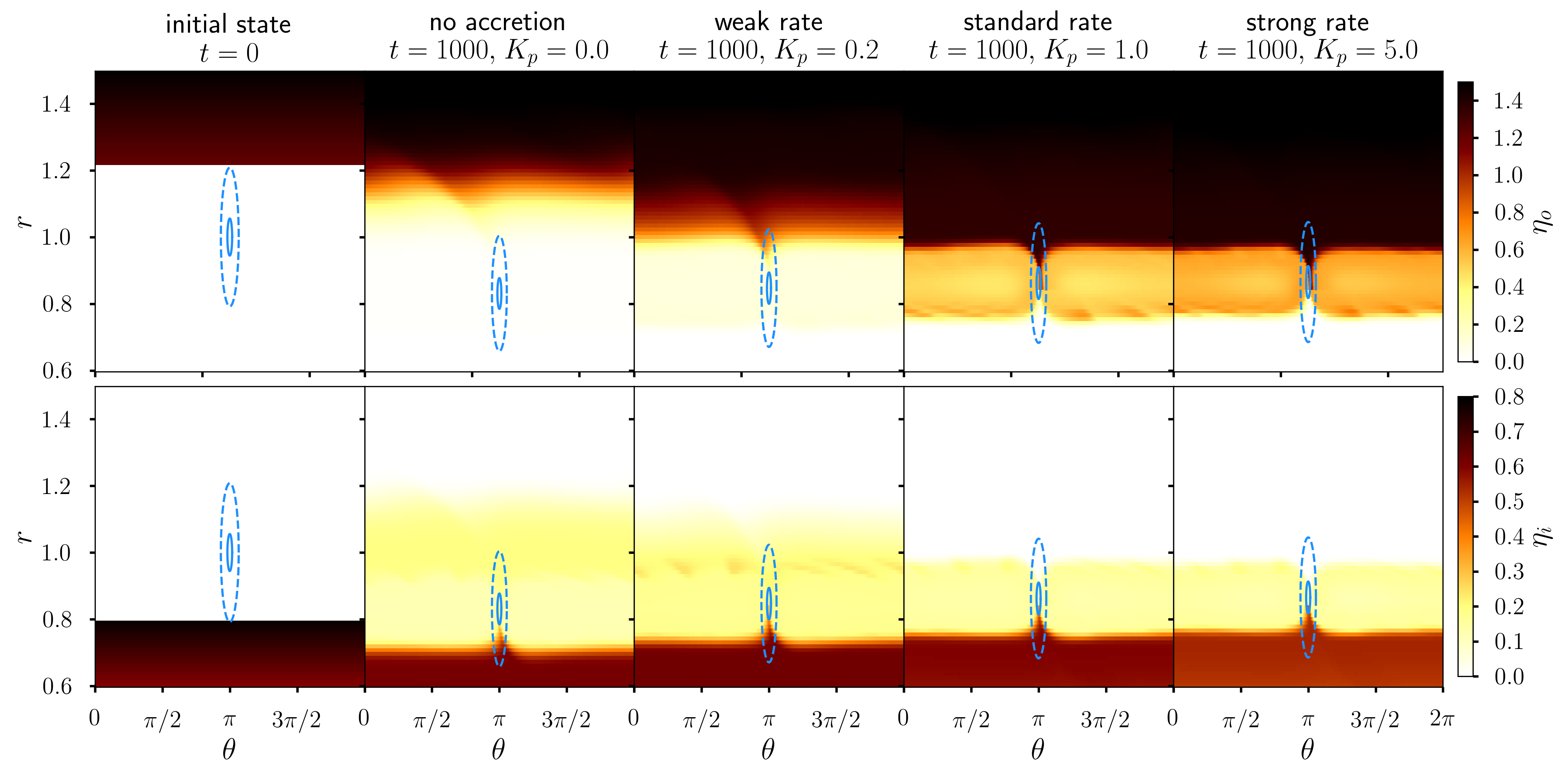}
  \caption{Evolution of passive tracers $\eta_{i/o}$ for simulation
    set {\tt A1}, following material originated in the
    outer\emph{(Top)}/inner\emph{(Bottom)} disk, in polar
    coordinates. \emph{(Left)} : initial state (white is $0$).
    \emph{(all but leftmost column)} : $1000$ orbits after release,
    for varying values of $K_p$. We emphasis that color scales are different
    across rows. For the sake of readability, the disk is here
    displayed with angular coordinates such that $\theta_p = \pi$ in
    every frame. Filled lines show the planet's feeding zone, $0.8R_H$
    in radius\,; dashed lines are $3R_H$ large in radius, encompassing
    a somewhat broader region than the typical HSR.  }
  \label{fig:tracers_A1}
\end{figure*}

\begin{figure*}
  \centering
  \includegraphics[width=\hsize]{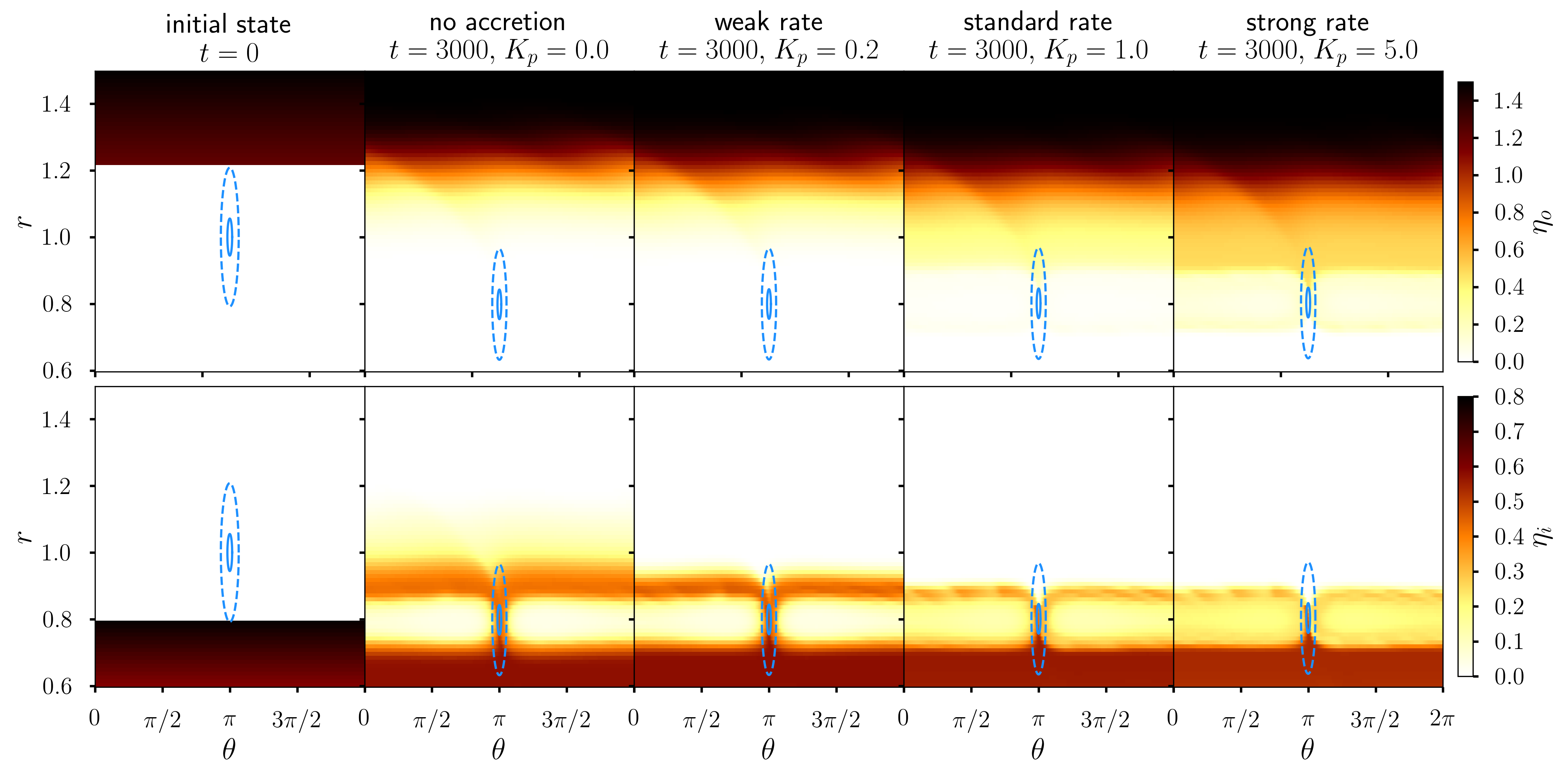}
  \caption{same as \Cref{fig:tracers_A1} for simulation set {\tt
      A2}. Snapshots are taken after a $\alpha_1/\alpha_2$ longer
    period.}
  \label{fig:tracers_A2}
\end{figure*}

The inner and the outer disks constitute our two gas reservoirs of
interest.
In order to keep track of the planet's \emph{relative} radial
displacement \emph{with respect to} those reservoirs, we use passive,
dimensionless, scalar tracers $\eta_{i/o}$.
Let us give a non-ambiguous and partly arbitrary definition of their
respective initial distributions.
We should be concerned about avoiding confusion with material
originated from the HorseShoe Region (HSR), whose width is given by
\citep{masset+2006} as $w_{\mathrm{HSR}}\approx 2.5\,R_H$.
Hence, a precautionary choice is to consider only material initially
distant of at least $3R_H$ from the planet's orbit.
Within this definition, we define $\eta_{i,o}(r,\theta,t=0) = r/r_0$,
for $r\gtrless r_p \pm 2.5R_H$, and $\eta_{i/o} = 0$ elsewhere.
Let us acknowledge that, as a tracer is advected along with the gas,
its value in a given cell becomes the mass weighted average of the
tracers that are found in the considered cell at the considered time.
Therefore, at $t>0$, we expect to find mixed material, displaying
values of the tracer that do not correspond to the initial value but
rather a weighted average of it.

\Cref{fig:tracers_A1,fig:tracers_A2} display, for simulation sets {\tt
  A1} and {\tt A2} respectively, the evolution of those tracers,
sampling over our parameter $K_p$ from $0.0$ to $5.0$.
Once more, following \citet{duffell+2014} and DK15, we find that, in
the non-accreting case, some gas is effectively transported from the
inner disk to the outer disk as migration proceeds.
Not only does the planet migrate \emph{with respect to} the medium, it
also actively ejects some material to larger orbits.
This still holds in the weakly accreting case $K_p=0.2$, although the
transport efficiency is being slightly decreased.
However, it is not so in the ``standard'' and ``strong'' accretion
cases ($K_p=1$, $K_p=5$), where the outward flow from the inner disk
is so efficiently blocked that the outer disk rushes into the planet's
vicinity.

Let us observe that while we successfully introduced accretion as a
means to prevent gap-crossing flows, the procedure also profoundly
modified the nature of the flow and added complexity to the picture.
Indeed, not only did we prevent inner disk material to transfer into
the outer disk, we also allowed the outer disk material to reach the
planet's feeding zone, hence causing depletion to happen in both
halves of the disk.
Additionally, we note that despite the \emph{origin} of gas at a given
radius being widely different depending on the accretion rate (see
\Cref{fig:tracers_A2}), at lower viscosity values ({\tt A2}) the
disk's structure stays almost self-similar whatever the planetary
accretion efficiency $K_p$, see \Cref{fig:profiles}.
The net effect on migration is unclear at this point and will be
discussed in \Cref{ssec:acc_mig}.
Furthermore, the transition from a naturally occurring gap-crossing
flow to a two-way accretion flow with increasing $K_p$ appears to be smooth.
Indeed, we see in the weakly accreting case that, as less material
flows from the inner to the outer disk, the latter immediately reacts
and follows the planet more closely than in the non-accreting
scenario.
From a quick comparison between \Cref{fig:tracers_A1} and
\Cref{fig:tracers_A2}'s respective ``weak rate'' panels, one can see
that the stronger the viscosity, the quicker this compensation
mechanism (normalizing time to the viscous timescale).
This is confirmed in stronger accretion cases $K_p=1.0,5.0$\,.
Hence, at least for high values of $\alpha$, planetary accretion can
not cut the otherwise existing outward\footnote{relative to the
  planet} gap-crossing flow without causing a more rapid inward flow
of the outer disk.

In short, we find that an accreting planet does not prevent gas from
entering its HSR, even though it can nullify gap-crossing
flows.
Let us now discuss the implications of accretion on the migration
speed.

\subsection{Impact of accretion on migration}
\label{ssec:acc_mig}
It is clear from our previous observations that the density of the
outer disk, hence the negative torque it yields on the planet should
be reduced by planetary accretion.
DK17 already noted that planetary accretion can reduce migration
speeds. Nonetheless, we stress that the positive contribution from the
inner disk can also diminish as the disk is being forcedly depleted.
Here we measure the effective migration rates against accretion 
efficiency and give further interpretation.

Because planetary accretion stops the gap-crossing gas flow, one may
expect it to consolidate the classical type~II migration scheme.
We recall that in this scheme, the migration rate is given by
$\dot{r_p}(r_p) = v_r(r=r_p)$, where $v_r$ is the viscous speed of the
unperturbed disk \Cref{eq:steady_vr}\,; i.e. the planet is assumed to
migrate with the drift rate of the gas.
Integrating \Cref{eq:steady_vr} we get the analytical evolution of
$r_p(t)$ for a planet migrating in a classical type~II fashion\,:
\begin{equation}
  \label{eq:analyticaltypeII}
  r_{p,\mathrm{II}}(t) = \left[r_p^{3/2}(t=0) - \frac{9}{4}\alpha h^2 \sqrt{Gm_*}t
    \right]^{2/3}.
\end{equation}

Time evolution of $r_p$ is shown on \Cref{fig:smaVSnorb} for various
accretion efficiencies ($K_p$), against this theoretical track, for
simulation sets {\tt A1} and {\tt A2} (1 panel per value of $\alpha$).
We observe that planetary accretion has significant impact only in the
highest viscosity case $\alpha=0.003$, where the migration rate can be
reduced below the theoretical rate (cases $K_p \ge 1$) after a few
$1000$ orbits in agreement with findings from DK17.
However, changes in migration are barely noticeable in the second case
where $\alpha=0.001$.
We have checked that for lower viscosities (simulations {\tt A3}, {\tt
  A4} and {\tt A5}), the migration speed is hardly impacted by
planetary accretion as well.
Increasing the accretion efficiency of the planet (thus effectively
cutting the flow of gas across the gap) does not allow the planet to
decreases its migration speed down to $v_r$ if $\alpha\leqslant
0.001$.

The role of accretion in the viscous case can be understood with
corresponding density profiles (\Cref{fig:profiles}).
For the {\tt A1} runs, the local disk's structure is being efficiently
affected by the accretion, decreasing both the inner and the outer
disk's density, hence their respective torques densities (as shown in
\Cref{fig:torques}).
In contrast, the planet can hardly make a dent in the density profile
in the lower viscosity cases.
This is consistent with that a lower viscosity implies both a wider
and deeper gap\,: as far less material is available in the vicinity of
the separatrices of the HSR, the influence of planetary accretion on
the dynamical evolution of the disk is hindered.

Except for high values of $\alpha$, we can conclude from our study
that planetary accretion is not of significant importance for the
migration of giant planets.
Actually, the fact that the curves in \Cref{fig:speeds} overlap while
there is no planetary accretion suggests that the latter is not
important in setting the proportionality between the migration rate
and the disk's viscosity.
The reason why planetary accretion plays a role in the migration speed
in the viscous \texttt{A1} case is not because it cuts the flow, but
because it perturbs strongly the density profile and broadens the gap.

\section{Conclusion}\label{sec:conclu}

To summarize, we have found in \Cref{sec:viscosity} that although
type~II migration speed is proportional to the gas' viscosity, it is
not driven by the radial inward drift of the gas.
In particular, we confirmed that the giant planet can actually migrate
faster than the gas drifts, and even in a stationary disk.
We concluded that what drives type~II migration is the imbalance
between the torques felt by the planet from the inner and the outer
disk, as pointed out by DK15.
However, the width and shape of the gap is not directly linked to
the viscosity, especially at low $\nu$ where the pressure effects
are dominant \citep{crida+2006}\,; hence, we do not expect this 
torque imbalance to be proportional to $\nu$, in contrast with 
the observed migration speed.
In \Cref{sec:accretion}, we have seen that gap-crossing flows are
actually negligible at low viscosity, and that cutting this small gas
flow with planetary accretion hardly impacts the migration speed.
Thus, the planet migrates faster than the disk drift, even when no gas
is exchanged between the inner and outer disks.
Gap-crossing flows can not be responsible for the observed fast
migration, in contrast with the of case type~III migration
\citep{masset-papaloizou2003}.

These two results allow to us draw a new, consistent picture of type~II
migration.
As a giant planet forms, it opens a gap by perturbing the gas profile
with the gravitational torque it exerts.
The gas reaches a new equilibrium profile on each side of the gap.
Nonetheless, the planet inside its gap feels a non-zero torque,
because the inner and the outer torques have no reason to balance out
\citep[as recently studied by][]{kanagawa+2018}.
Thus, the planet has to migrate inwards.
As it does so, some gas may cross the gap from the separatrix of the
HSR, although this is not enough to restore the initial gap profile in
the frame of the planet if the viscosity is low ad the
  gap is wide (regardless of whether the planet accretes or not).
Therefore, the density distribution has to adapt to the new position
of the planet\footnote{i.e. the bump in the surface density produced
  at the inner edge of the gap has to be redistributed over the inner
  disk while the outer disk has to spread down to the new gap's outer
  edge.}, and this is done over a viscous time.
Once the gas is again at equilibrium with the planet, the planet is
not in an equilibrium inside the gap anymore, and we are back to the
initial situation.

In this scheme, the planet may well migrate faster than the gas
drifts, because it is pushed by a torque that has no connection with
the drift of an unperturbed disk.
But because the gap-crossing flow is negligible, the planet must
migrate at a rate proportional to viscosity, otherwise it would pile
gas up in the inner disk and leave a depleted outer disk behind,
eventually halting its migration.

Although the final result (migration speed of gap-opening planets is
proportional to the viscosity) is in line with the standard picture of
type~II migration, this new scheme is conceptually revolutionary in
our understanding of this phenomenon, and allows us to reconcile all the
puzzling observations that have been made recently, questioning the
standard picture.
Additionally, it confirms that even if some gas may cross the gap, a
giant planet in a low viscosity disk should migrate slowly.
In this frame, the abundance of warm Jupiters, who did not migrate all
the way towards their star, may suggest that most protoplanetary disks
have a low effective viscosity in the planet-forming region.

\begin{acknowledgements}

We acknowledge support by the French ANR, project number
ANR-13-BS05-0003-01 projet MOJO. (\textbf{M}odeling the
\textbf{O}rigin of \textbf{JO}vian planets).
C.M.T Robert, H. Méheut and A. Crida acknowledge funding from ANR number ANR-16-CE31-0013 (Planet-Forming-Disks).
HPC resources from GENCI [IDRIS]
(Grant 2017, [i2017047233]) and from "Mesocentre SIGAMM", hosted by Observatoire de la C\^ote d'Azur were used.

C.M.T Robert and E. Lega wish to thank Alain Miniussi for his valuable help in maintaining and developing the code base used in this work.
Figures in this paper were produced with {\tt matplotlib} \citep{Hunter:2007}.
The authors wish to thank the anonymous referee for their valuable observations and help in making this paper clearer.
\end{acknowledgements}

\afterpage{
\begin{figure*}
\includegraphics[width=\hsize]{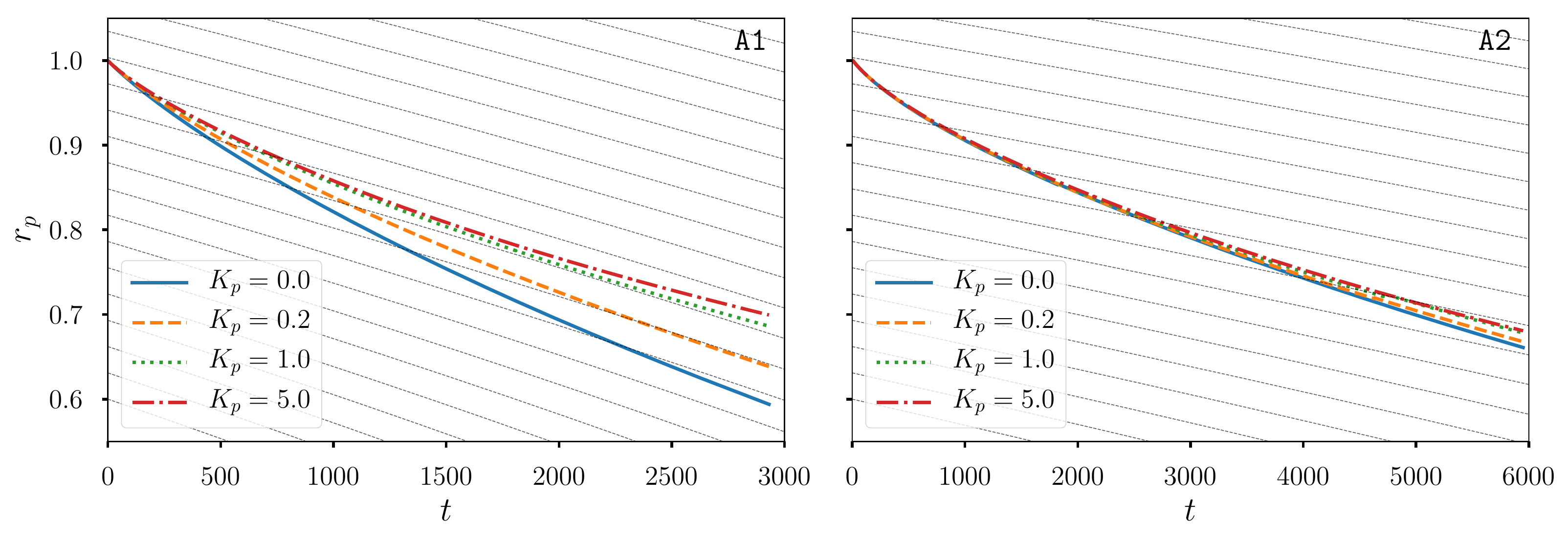}
  \caption{Radial position of the planet $r_p$ VS time, for varying
    accretion rates. Panels correspond to simulation sets {\tt
      A1}(\emph{Left}) and {\tt A2}(\emph{Right}). The plot includes
    analytical migration tracks (\Cref{eq:analyticaltypeII}) with
    varying time offsets as guidelines.}
  \label{fig:smaVSnorb}
\end{figure*}

\begin{figure*}
  \includegraphics[width=\hsize]{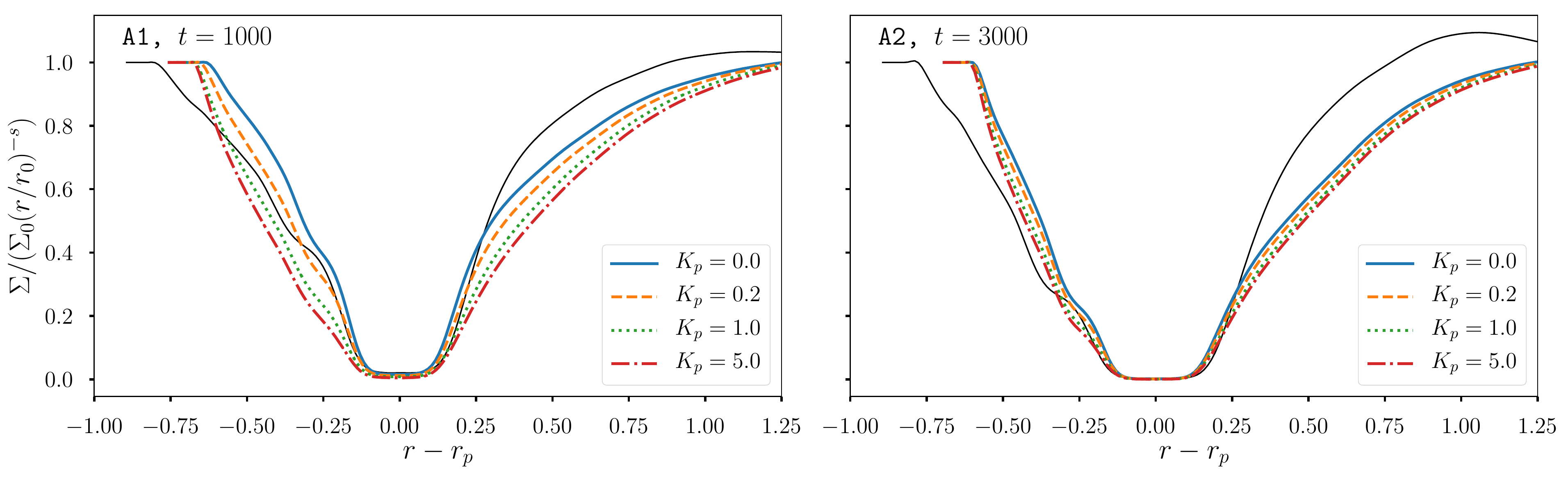}
  \caption{Density profiles snapshots at fixed time with varying
    accretion efficiency for simulation sets {\tt A1} and {\tt
      A2}. Black thin lines indicate the initial states (same as
    \Cref{fig:init_density}).  }
  \label{fig:profiles}
\end{figure*}

\begin{figure*}
  \centering \includegraphics[width=\hsize]{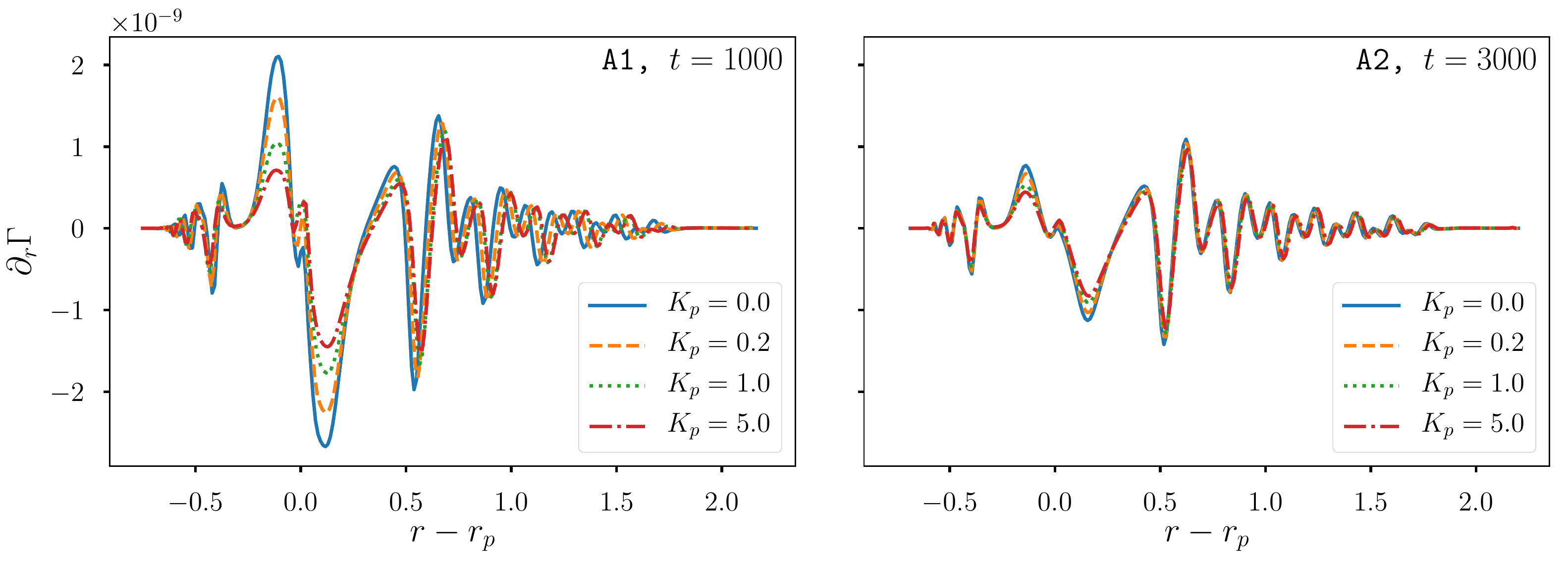}
  \caption{Radial torque densities For simulation sets {\tt A1} and {\tt A2}.}
  \label{fig:torques}
\end{figure*}
\clearpage
}

\bibliographystyle{aa}
\bibliography{references}

\begin{appendix}
\section{Disk relaxation\,: equilibrium of torques on a fixed-orbit planet}
\label{app:ics}
\begin{figure}[h]
  \includegraphics[width=\hsize]{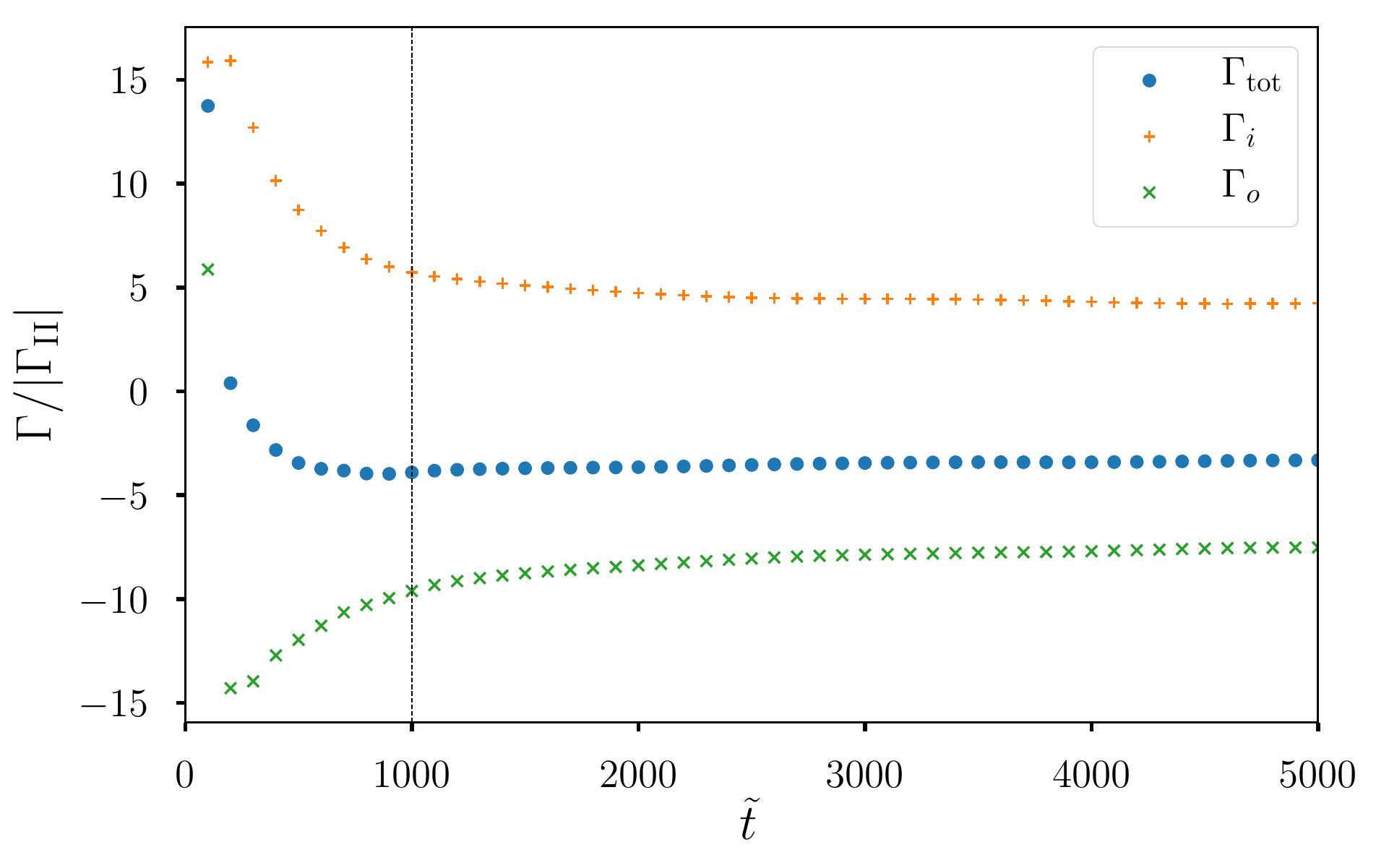}
  \caption{Evolution of the gravitational torque $\Gamma_\mathrm{tot}$
    acting on the planet during \emph{introduction} and \emph{relaxation}
    ($\tilde{t}$ here denotes time during those stages).
    $\Gamma_i$ is the contribution of the inner disk (up to
    $r=r_p=1$), $\Gamma_o$ of the outer disk, and $\Gamma_\mathrm{tot} =
    \Gamma_i+\Gamma_o$.
    A vertical line indicates the end of the planet's \emph{introduction} and starting date of the disk's \emph{relaxation}.
    \label{fig:torque_convergence}
    }
\end{figure}

Here we give details on the \emph{relaxation} of the disk towards
initial conditions for planetary migration.
On \cref{fig:torque_convergence}, the total gravitational torque
$\Gamma_\mathrm{tot}$ that the disk exerts on the planet is plotted
alongside with its inner and outer components.

While the one-sided torques decrease in absolute value as the gap
opens, the total torque remains nearly constant.
We measure its final value as only a fraction ($\sim0.12$) of the differential
Lindblad torque
\begin{equation}\label{eq:lindbladtorque}
  \Gamma_0 = - \Sigma_p r_p^4 \Omega_p^2 (q/h)^2\,,
\end{equation}
where $\Sigma_p$ is the \emph{unperturbed} surface density at
planetary semi-major axis, $r_p$\,.
This is expected for a gap-opening planet.
For lack of an existing normalization factor for type~II migration, let us derive one here.
The viscous torque exerted by inner material $r<R$ onto its complementary part $r>R$
in an unperturbed disk may be expressed as
\begin{equation}
  \Gamma_\nu (R) = 3\pi\nu\Sigma R^2\Omega(R)
  = 3\pi\alpha H^2\Omega^2\Sigma R^2\,.
\end{equation}
Hence, the expected value the total torque in classical type~II can be estimated as the
differential viscous torque that a gas annulus occupying the gap would feel as
\begin{equation}
  \label{eq:gammaref}
  \Gamma_\mathrm{II} = 3\pi\Sigma_0\alpha h^2 G m_* \left[
    (r_p-w_\mathrm{HSR})^{1-s} - (r_p+w_\mathrm{HSR})^{1-s} \right]\,.
\end{equation}
where $w_\mathrm{HSR}$ is the width of the horse-shoe region, estimated as $2.5\:R_H$\,.
We observe that $\Gamma_\mathrm{tot}$ converges towards $\sim
4\;\Gamma_\mathrm{II}$.
This discrepancy has been tackled by \citet{duermann-kley2015} who
showed that in fact, the final value of this torque is primarily
determined by the disk mass has little to do with the disk accretion
rate $\dot{M}$, or equivalently its viscous torque $T_\nu \propto \nu \propto \dot{M}$.
Furthermore, they showed that even-though the inner and outer disks
reach stationary density profiles, the planet is not in an
equilibrium position between them.
Indeed, a stationary-state disk embedding a fixed-orbit planet does
not yield the expected type~II torque but rather a significantly stronger one. 

\section{Design of boundary conditions}
\label{app:bcs}

\begin{figure}[!h]
  \centering
  \includegraphics[width=0.7\columnwidth]{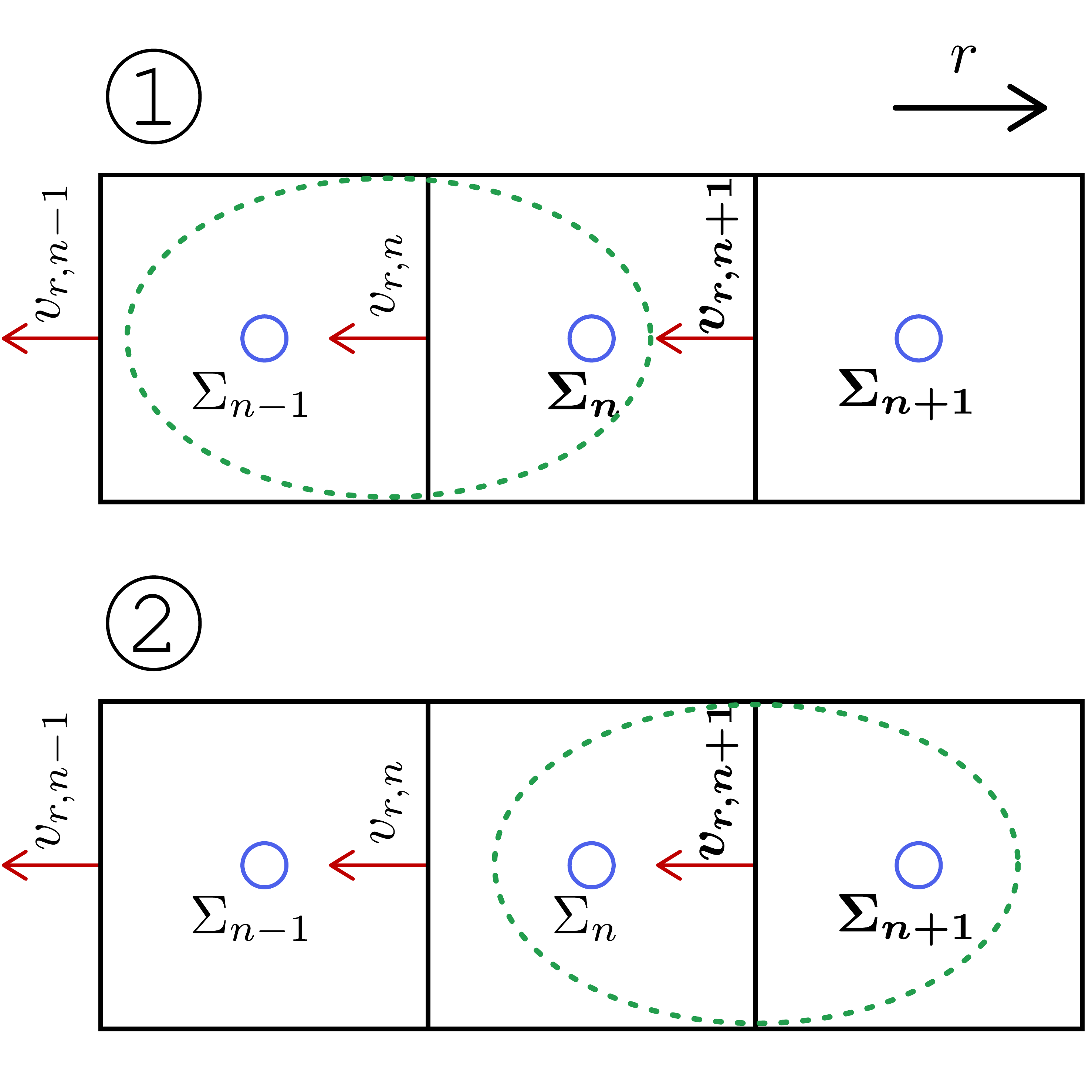}
  \caption{Two possible designs for fixed boundary conditions in a 1D
    grid where radial velocities are defined at inner cell edge (red arrows) and
    density is center-defined (blue circles).
    Boldface is used to indicate fixed quantities.
    The net mass inflow is determined as a combination of quantities inside of a green dashed ellipse, which always contain both fixed and free values.
    The last cell harbouring free quantities is indexed by $n$.
  }
  \label{fig:fluxscheme}  
\end{figure}

Here we provide more insight on how the boundary conditions
  were selected and acknowledge the consequences of this choice.
An obvious choice for boundaries would consist in simply fixing the
values of velocities and surface density to their initial state.
In principle, this design would allow for conservation of both the
total mass in the simulation domain and the initial flow $\dot{M}$
throughout the relaxation.
However, this does not hold within the staggered grid scheme used in \texttt{Fargo}.
Indeed, radial velocities are defined at lower edge of cells while
density is center-defined.
While this is a convenient design for numerical integration of
hydrodynamics equations, it makes it impossible to keep the
density, velocities and the net mass flux entering the simulation
domain constant all at once.
This is because the net mass entering the simulation at the outer edge
is effectively defined as a combination of fixed and free numbers.
\Cref{fig:fluxscheme} illustrates this point, and define the two equally
problematic possible designs.
A solution to this issue would be to correct the imposed value for
$\Sigma_n$ (case 1) or $v_{r,n+1}$ (case 2) at each time step to achieve the
desired value in $\dot{M}$.
In practice this proved less than efficient, as $\dot{M}(r)$ is very
sensitive to even the most subtle spatial or time variation,
convergence was not achieved in practical times.

Although a steady-state is characterized
  by uniformity in accretion rate $\dot{M}(r)$, there is in fact no
  strong argument in favor of enforcing its initial value at the
  grid's edges.
Indeed, DK15 already clarified that the migration rate is not bound to
the unperturbed $\dot{M}$.
Furthermore, as previously observed, enforcing a constant mass flow through
both radial edges would cause the total mass, dictated solely by $s$
and the radial limits of the grid,
$(r_\mathrm{min}\,,r_\mathrm{max})$, to be constant.
In this case, one would obtain qualitatively different results only by
enlarging the simulation domain, corresponding to different physical
scenarios.
For instance, strong pressure bumps may be found at the edges of the
gap for a narrow enough grid, corresponding to a fast planet-formation
scenario where a jovian mass would agglomerate over the course of a few
100 orbits.

For these reasons, we chose to relax the assumption that the
  final uniform value in $\dot{M}$ should be exactly equal to that of
  the initial, analytical state described in \Cref{ssec:model}, and
  instead favor the fact that the disk should stay close to
  unperturbed far from the planet.
As explained in \Cref{ssec:bc}, we extend the notion of ``boundaries''
to broad radial domains where perturbations with respect to the
initial state are consistently damped out.
This artificial process is applied to velocities (both radial and
azimuthal) as well as to surface density.
A direct consequence of this is that mass is no longer being conserved
in wave-killing regions, so that desired uniformity in $\dot{M}(r)$ is
only relevant outside those regions.
Additionally, this damping process prevents reflections of sound
waves, such as spiral wakes, as is its original purpose
\citep{Val-Borro2006}.

\begin{figure}[!h]
  \centering
  \includegraphics[width=\columnwidth]{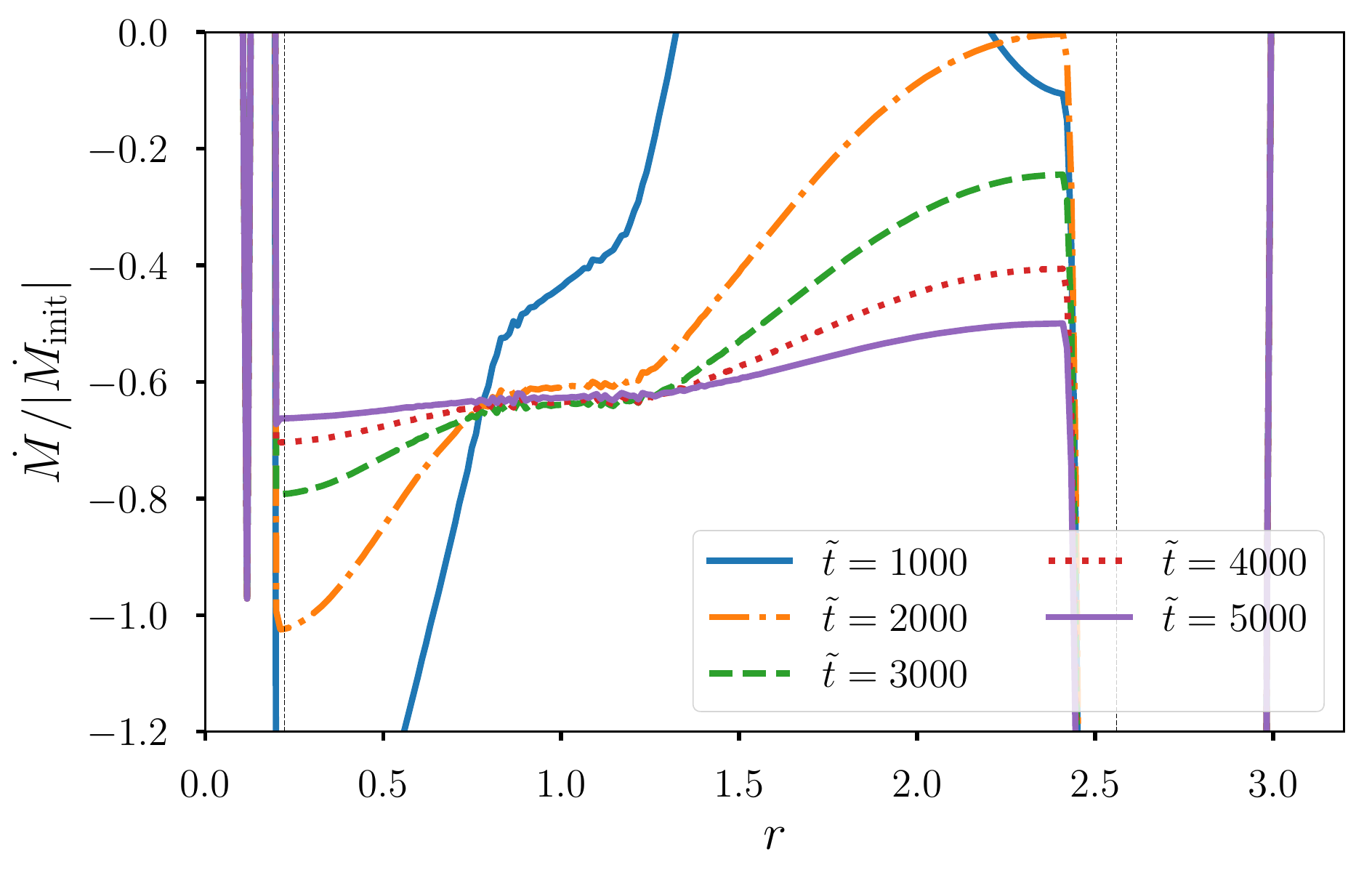}
  \caption{Relaxation of the azimuthally averaged radial flow, case
    \texttt{A1}.  $\tilde{t}$ denotes time during \emph{introduction}
    and \emph{relaxation} stages ($\tilde{t} = t +5000$).  Limits of
    the domain of interest (excluding wave-killing zones) are shown as
    dashed lines. $\dot{M}_\mathrm{init}$ denotes the uniform value in
    the analytic state of the unperturbed disk described in
    \Cref{sec:setup}.  628 snapshots covering one orbit were averaged
    to obtain each line on this figure.  }
  \label{fig:flow_convergence}
\end{figure}

\Cref{fig:flow_convergence} shows the evolution of the radial flow profile obtained in the case \texttt{A1}. Although not perfectly converged, at $\tilde{t} = 5000$, we settled for what we considered an acceptable level of uniformity to save computational time.

\end{appendix}

\end{document}